\documentclass[aip,pop,reprint]{revtex4-1}
\usepackage{graphicx}
\usepackage{amsmath}
\begin{document}

\title{Ultracold Plasma Expansion as a Function of Charge Neutrality}
\author{Craig Witte and Jacob L. Roberts} 
\address{Colorado State University, Fort Collins CO, 80523}
\begin{abstract}
Ultracold plasmas (UCPs) are created under conditions of near but not perfect neutrality. In the limit of zero electron temperature, electron screening results in non-neutrality manifesting itself as an interior region of the UCP with both electrons and ions and an exterior region composed primarily of ions. The interior region is the region of the most scientific interest for 2-component ultracold plasma physics. This work presents a theoretical model through which the time evolution of non-neutral UCPs is calculated. Despite Debye screening lengths much smaller than the characteristic plasma spatial size, model calculations predict that the expansion rate and the electron temperature of the UCP interior is sensitive to the neutrality of the UCP. The predicted UCP dependence on neutrality has implications for the correct measurement of several UCP properties, such as electron temperature, and a proper understanding of evaporative cooling of the electrons in the UCP.
\end{abstract}
\maketitle

\section{Introduction}
Ultracold plasmas (UCPs)  offer the opportunity to study plasma physics within a unique range of plasma parameters \cite{killian1999}. Simple estimates from UCP initial ionization conditions\cite{killian1999,killian2001,gupta2007}  suggest that experiments with the electrons and/or the ions deep in the strongly-coupled regime\cite{strong_coupling_review,Bannasch2013,Cummings2005}  should be able to be performed. This simple estimation is incomplete, however, as there are several heating mechanisms such as disorder-induced heating\cite{killian2001,gupta2007}, three-body recombination\cite{Denning2009}, and continuum lowering\cite{Hahn2001}  that raise both the ion and/or electron temperatures quickly and thus reduce the degree of strong-coupling shortly after UCP formation.  In principle, this heat could be subsequently removed and the degree of strong-coupling increased if cooling occurs in the UCP after formation. After formation the UCP expands due to the thermal pressure of the electrons, and so adiabatic cooling will lower the electron temperature.  Additionally, evaporative cooling can also cool the electrons\cite{coolingrev}. Recent work\cite{Wilson2013}  has shown that the influence of evaporative cooling increases as the density of a UCP decreases. While that work examined the cooling from the evaporation that inevitably occurs during the expansion of a UCP, forced evaporative cooling should be even more effective and can be implemented by using external electric fields to deliberately increase the evaporation rate from the UCP\cite{Roberts2004}.

One consequence of such forced evaporative cooling is a reduction in the charge neutrality of the UCP.  Therefore, to successfully interpret experiments investigating forced evaporative cooling in UCPs, the influence of reduced charge neutrality on UCP behavior, particularly the UCP expansion rate and subsequent electron cooling, needs to be studied and quantified.  This is the subject of the work presented here. Determinations of electron temperature during UCP expansion and forced evaporation experiments performed in the absence of this analysis have the potential to give misleading results, depending on the experimental techniques used. This is true not only of forced evaporation cooling experiments, but of any experiment where the UCP neutrality is varied that involves UCP expansion, electron temperature, or electron confinement as relevant parameters.

As part of the UCP formation process, an initially neutral ionized gas acquires a net positive charge as some electrons escape the UCP\cite{killian1999}. The net positive charge creates a potential well, confining most of the electrons inside the UCP\cite{simple}. For sufficiently cold electrons, the UCP can be thought of having two distinct regions. The interior is characterized by the presence of significant ion and electron densities, while the exterior is predominantly composed of ions.  Insofar as the UCP is a two-component plasma, the interior region is the main region of interest.  The inner region is where oscillations in the electron component are located, where three-body recombination is significant, and where electron response to external fields will be centered.

For typical quasi-neutral UCP conditions, the post-formation expansion is driven primarily by the thermal pressure of interior electrons and not by Coulomb forces that arise from the overall charge imbalance between the number of ions and electrons.  At first glance, it would seem that any increase in charge imbalance would result in an increase in the total UCP expansion rate due to an increase in electrostatic forces. However, it remains unclear whether the expansion of the UCP interior would be impacted by additional Coloumb driven expansion.  Gauss' Law and the spherical symmetry of UCPs imply no additional electric field in the interior of the UCP from an increase in the number of ions in the exterior region.  Moreover, any electric fields would be screened by the electrons in the UCP.

Despite screening and the spherical symmetry considerations, however, there are two possible ways that a change in charge neutrality could affect not only the total UCP expansion rate but the expansion rate of the interior region of the UCP as well.  Changes in the exterior charge distribution will alter the expansion rate of that part of the UCP and that in turn will influence the expansion rate at the boundary of the interior region.  A change in the rate of expansion at the interior region boundary can lead to an overall change in the interior region expansion rate as a whole.  Furthermore, a change in charge imbalance will alter the UCP potential depth.  This in turn influences the confined electron spatial density distribution, possibly affecting interior expansion. The work in this article theoretically investigates UCP expansion and electron cooling as a function of the neutrality of the UCP.

Our work synthesizes previous theoretical models that have been developed for describing UCP expansion\cite{simple,Robicheaux2003}.   In general, UCP near-neutrality is assumed in previous published work\cite{simple,gupta2007,Bergeson2011,Fletcher2007}  and this results in thermal energy being the primary driver of UCP expansion. Our primary contribution is to extend calculations such as in Ref. \cite{gupta2007,simple,Kuzmin2002}  to treat circumstances of extremely reduced charge  neutrality.

It was found that increased charge  imbalances resulted in a significantly increased rate of exterior region ion expansion. This additional ion exapansion is large enough to be detectable by ion absorption spectroscopy commonly used in UCP experiments\cite{killian2005,Lyon2011}. This increased rate of exterior expansion is found to lead in turn to additional cooling of the UCP electrons via adiabatic expansion cooling. The relative impact of this additional cooling becomes greater as the initial UCP electron temperature is decreased. The increased cooling in the interior region reduced the magnitude of thermal pressure, causing a small, but detectable slowing of the interior expansion well inside the boundary. Thus, increases in charge imbalance are associated with changes in the spatial shape of the UCP ion distribution. From the perspective of trying to remove energy from the electron component to reduce its temperature, such additional cooling from reduced neutrality is beneficial.  The mechanism for this additional electron temperature reduction as a function of different UCP parameters is treated in the following sections.  

\section{UCP Expansion and Electron Temperature Model}
Our model calculates the path of UCP ions over the course of the UCP expansion.  Spherical symmetry is assumed for both the ion and the electron spatial distriution.  At each point in time the electrons are assumed to be in thermal equilibrium, modified slightly from the Maxwell-Boltzmann distribution as discussed below.  This assumption of thermal equilibrium requires particular average electric fields to be present that confine the electrons, and these fields drive the ions outward in expansion.  The expansion of the ions can thus be modeled via computing these electric fields and their effect on UCP ions.  Energy conservation was used to determine the change of electron temperature with time.  Different UCP initial parameters can be altered (e.g. electron temperature, UCP size and density, neutrality) to determine their influence on the UCP expansion
 
The ions in the UCP are treated as zero temperature particles and assumed to be initially distributed with a Gaussian density distribution described by $\frac{ N}{(2\pi\sigma_0^2)^{3/2}}e^{-r^2/2\sigma_0^{2}}$, where N is the number of ions, $\sigma_0$ is the spatial scale, and r is the distance from the UCP center. To track the change in the ion density distribution with time, the ions are assigned to a series of thin concentric spheres (shells). To avoid finite number issues 10,000 shells are used. These shells are extended over a distance of 5$\sigma_0$, after which the distribution is truncated.  When necessary, the ion shells were smoothed into a continuous density function by utilizing a series of interpolating polynomials. This is  necessary, for instance, in the determination of the electron density distribution.

 UCP electrons are described by an energy distribution equal to a Maxwell-Boltzmann distribution at lower energies but tending smoothly to zero at an upper cut-off energy corresponding to the depth of the electron confining potential\cite{king}. This is known as the King model of the electron thermal equilibrium distribution as decribed in Ref. \cite{king} . This model has been compared to more sophisticated Monte Carlo methods describing UCP electrons, and has exhibited good agreement with the more sophisticated Monte Carlo technique.  The need for a cutoff exists because in general the net potential energy from the ions and electrons tends to a finite value far from the UCP.  For a purely Maxwell-Boltzmann distribution this would imply a non-zero density of electrons everywhere outside the UCP, leading to a predicted infinite number of electrons. This non-physical result necessitates a truncation of the distribution.   This truncation point is labeled as $R_0$. We set $R_0$ equal to $6\sigma_0$ in our computations, which is reasonable based on typical extraction fields applied in UCP experiments\cite{Wilson2013,chen2013}.
 
To compute the electron distribution, the technique outlined in \cite{king} is followed.  Poisson's equation can be expressed as:
\begin{equation}
\frac{1}{R}\frac{d^2}{dR^2}RW(R)=-\frac{q\sigma_0^2}{k_BT\epsilon_o}(\rho_e(W(R))+\rho_I(R))
 \end{equation}
Here,  $\rho_I$ and $\rho_e$  are the ion and electron charge density respectively. W is a scaled electrostatic potential, such that $W=e(\phi(R)-\phi(R_0))/kT$, where $\phi$ is the electric potential. It is important to note that $\rho_e$ is an explicit function of $W$ (see \cite{king}), while $\rho_I$ is completely independant of $W$. $W$ is calculated self-consistantly by using a numerical relaxation technique, and with a known $W$ the electron charge density and the net electric field in the UCP can be computed explicitly.

We note that unlike a Maxwell-Boltzmann distribution, the quantity $T$ in the King distribution is not truly a single temperature. Instead, it acts as a model parameter. A position-dependent effective temperature can be defined such that $\frac{1}{2}m<v^2>=\frac{3}{2}k_BT_{eff}$ at a certain radius\cite{king}. This fact has consequences for interpreting the electron cooling during expansion properly, as discussed below.  The local electron temperatures are still exclusively a function of $T$ and $W$ though. Appendix A contains a more detailed discussion of the relationship between T and temperature.  In addition, we have included a description of all of the equations used in the model in appendix A as well.

With the average electric fields determined using the self-consistently computed electron distribution, the ion expansion was computed. To do this, the ion expansion was  broken up into a series of timesteps. At each time step, the electric field was calculated from the derivative of $W$. From these fields the force on each shell was determined. Each shell was accelerated for the duration of the time step in accordance with the force it experienced and its position was advanced by the amount determined by its velocity. New shell positions and velocities were then recorded for use in the next timestep.  Ion-ion collisions are not expected to play a significant role over the timescale of the calculations performed\cite{Spitzer}, and ion shells were allowed to pass through each other.

As mentioned above, UCP electron cooling rates are determined by assuming energy conservation.  The total energy is a sum of the electrostatic energy of the total charge distribution, the ion kinetic energy, and the electron kinetic energy (as specified locally by $T_{eff}$).  The total energy ($U$) can be expressed as:
\begin{multline}
U=2\pi\epsilon_0\int_0^{r_0}r^2E(r)^2dr+\frac{1}{2}Nm_i<v_i^2>\\
+6\pi k_b\int_0^{r_0}r^2n_e(r)T_{eff}(r)dr
 \end{multline}
where $\epsilon_0$ is the permittivity of free space, $E(r)$ is the electric field,  $N$ is the number of ions, $m_i$ is the mass of an ion, $v_i$ is the velocity of the ions, and $n_e$ is the number density of electrons. In all of our calculations, the number of electrons were held constant, and thus $U$ is held fixed.  As the UCP expands, ions are accelerated and their kinetic energy increases.  Expansion leads to a decrease in charge density, which causes a decrease in the electrostatic energy. Finally, electrons transfer thermal energy to ion kinetic energy, leading to cooler electrons.  Changes in ion kinetic energy and electrostatic energy were calculated explicitly. Decreases in electron thermal energy followed from energy conservation. $T$ was adjusted throughout the modeled expansion to produce the appropriate decrease in electron thermal energy required to maintain energy conservation.

The model, as currently constructed, does not account for any evaporation after the initial charge imbalance is specified. Not only does this simplify the needed calculations, it also isolates the effect of the charge imbalance from other effects during the UCP expansion.  Also, for many experimentally achievable UCP initial conditions additional evaporation would be negligible after a period of forced evaporation since in that case low electron temperatures and high charge imbalances would result. In future work we plan to incorporate additional evaporative cooling into the model in conjunction with experimental measurements of the evaporation rate\cite{antiproton}.
\section{Results}
The effect of charge imbalance on the rate of UCP expansion can be illustrated by comparing the results of a pair of UCP simulations. We quantify the degree of charge imbalance by the parameter $\delta=(N-N_e)/N$ where $N$ is the number of ions and $N_e$ is the number of electrons. For the simulations serving as the illustration of typical results, charge imbalances of $\delta$=0.1 and $\delta$=0.4 are  used.  Both simulations had initial parameters of $\sigma_0=3.75\cdot10^{-4}m$, $N=3.0\cdot10^5$ and $T=85K$.  For the ion mass in these calculations, we used \textsuperscript{85}Rb. Time steps were set to be 50ns, and the simulations ran for 85 steps (4.25$\mu s$).

The first noticeable differences between the two cases was visible in their electric field vs. position at time t=0. In both simulation runs there was a center region where the electric field matches that computed by using a simple model that assumes a neutral plasma\cite{simple}.  Far away from the UCP, the electric field behaves as expected and matches that of a point charge at the origin whose magnitude is equal to the charge imbalance. However, the field in between these two regions is strongly influenced by the UCP charge imbalance. For the $\delta$=0.4 simulation, this middle region has a pronounced electric field peak whose magnitude is much larger than the electric field associated with the neutral plasma approximation. In contrast, the electric field in the more neutral case follows the approximate field with only slight positive deviations until the 'edge' of the UCP is reached.  At that point, the electric field in the $\delta=0.1$ case slowly declines to its 1/r asymptotic behavior.

\begin{figure}
\includegraphics{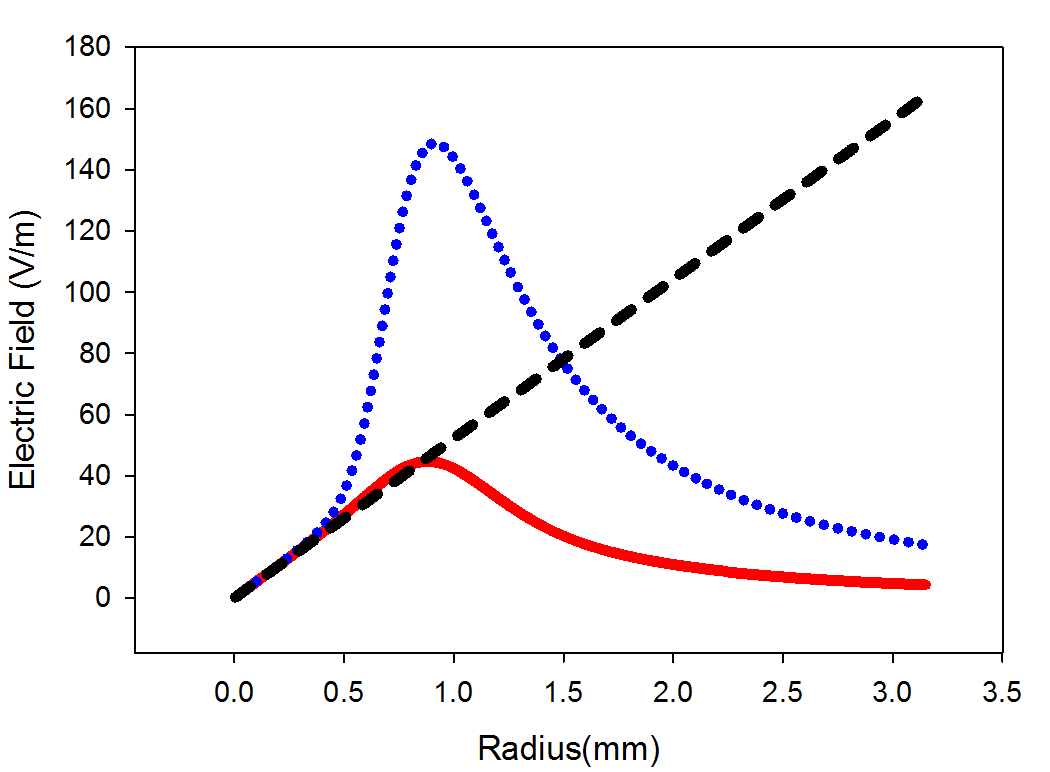}
\caption{ Initial electric field vs. position for two \textsuperscript{85}Rb UCPs with different charge imbalances. The solid line is a $\delta=0.1$ plasma, and the dotted line is a $\delta=0.4$ plasma. Both UCPs have initial parameters $T$=85K, $\sigma_0$=3.75x10\textsuperscript{-4}m, and $N$=3x10\textsuperscript{5}. The dashed line is from a simple model for a 85K neutral plasma as described in the main text. The electric fields at the center of the plasmas mirror the predictions of the simple model well. However, once outside the center, the $\delta$=0.4 field increases substantially while the $\delta$=0.1 field increases only slightly.}
\end{figure}

This  difference in electric field structure had a significant impact on the expansion of the UCP. The $\delta$=0.1 simulation’s expansion was approximately uniform, maintaining an approximately Gaussian shape throughout the simulation. The $\delta$=0.4 simulation, however, showed a much larger degree of exterior expansion, as can be seen in Fig. 2. The ions in the aforementioned middle region undergo a much larger acceleration than the rest of the ions outside the middle region, and so the ions initially residing farther out in the UCP exterior ultimately had lower velocity than the ions that started in the region where the electric field is the largest. The difference in velocities led to interior ions eventually catching up to exterior ions and forming a large ion spike at the edge of the UCP, as originaly theoretically predicted in Ref. \cite{Robicheaux2003} and and also observed theoretically in Ref. \cite{ionspike}. The extra expansion also led to a larger overall UCP spatial size as compared to the $\delta=0.1$ case. Our predictions confirmed the naive expectation that larger charge imbalances lead to higher rates of overall UCP expansion despite electron screening in the interior of the UCP. 

\begin{figure}
\begin{center}
\includegraphics{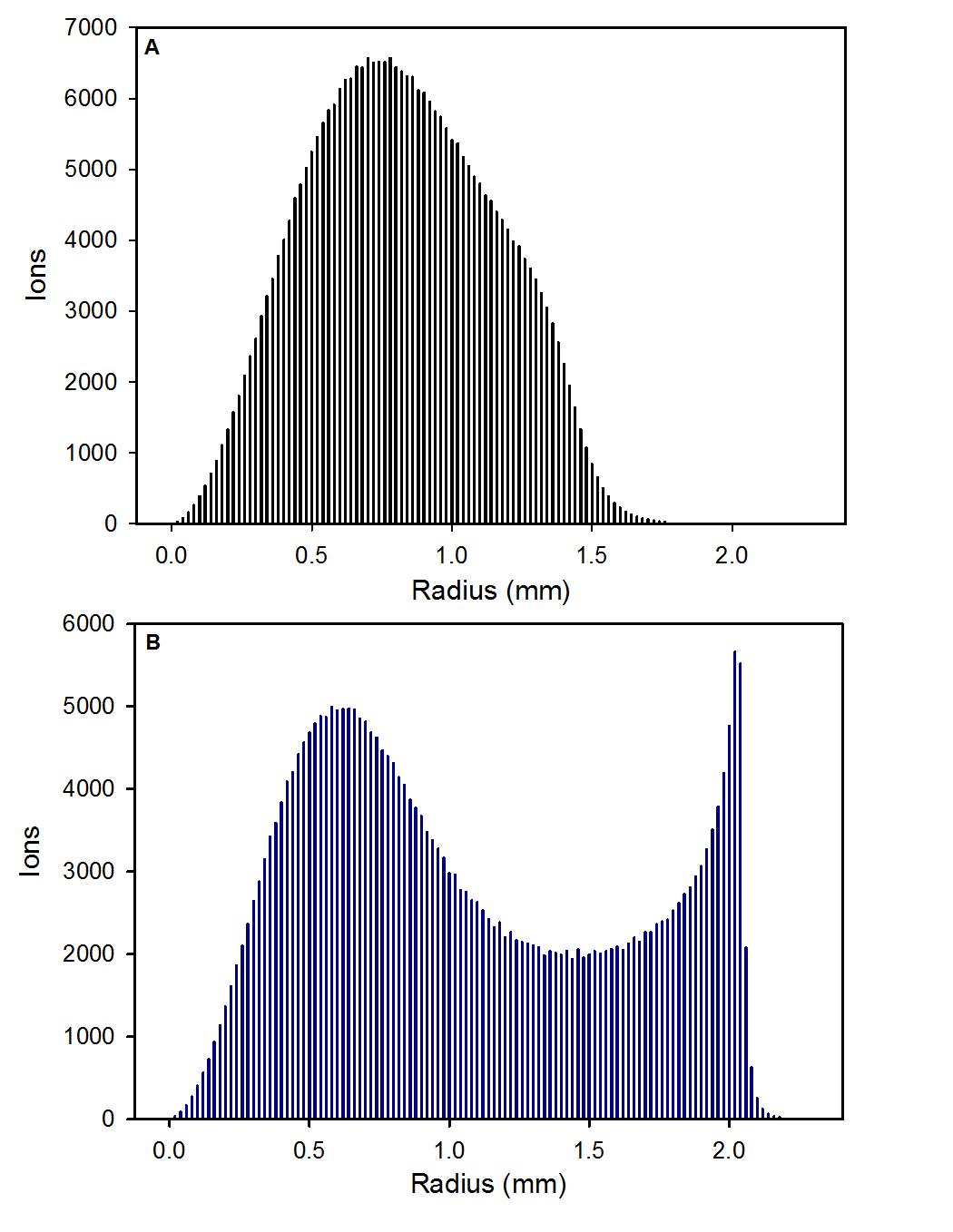}
\end{center}
\caption{A) An example of the Ion distribution of a plasma after expanding for 4$\mu$s. This particular plasma had initial parameters $T$=85K, $\sigma_0=3.75x10^{-4}$m, $N=3x10^5$, and $\delta=0.1$.  The ion distributions remains mostly Gaussian, but a small bulge is detectable at the  edge. B) A second plasma, identical to the plasma in (A), except that fvor this case $\delta=0.4$.  The increase in charge imbalance produced substantially more expansion on the exterior and resulted in the formation of a so-called ion spike on the edge.}
\end{figure}

While the charge imbalance has a large and visible impact on the exterior of the plasma, the variation of the expansion of the interior is shielded to some extent from effects associated with changes in the charge imbalance. This is reasonable given the screening properties of a plasma. The radius of a sphere that encloses 90\% of the electrons in a $\sigma_0$=3.75x10\textsuperscript{-4}m, $T$=85K, $N$=3.0x10\textsuperscript{5}, and $\delta=0.1$ UCP is approximately 900$\mu$m. Electron densities at the edge of this sphere are approximately 2.5x10\textsuperscript{13}m\textsuperscript{-3} which corresponds to a Debye length of 127$\mu$m. Since the UCP is larger in spatial extent than the Debye length, it is reasonable to expect that edge perturbations would have little effect on the center. The UCP electric field reinforces this view. At small values of $R$, the electric field does not change as a function of charge imbalance.

\begin{figure}
\includegraphics{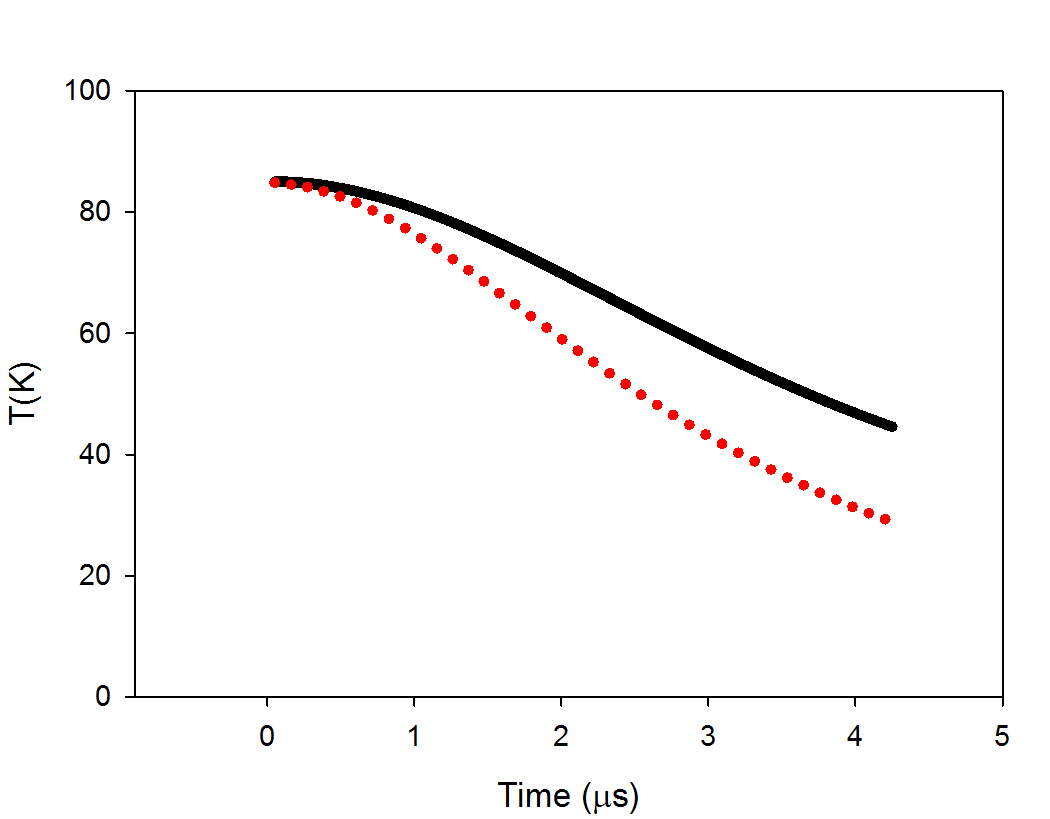}
\caption{ Plots of $T$ vs time for two plasmas with different charge imbalances. The solid line is a $\delta$=0.1 plasma, and the dotted line is a $\delta$=0.4 plasma. Both plasma’s have initial parameters $T$=85K,$\sigma_0$=3.75x10\textsuperscript{-4}m, and $N$=3x10\textsuperscript{5}.}
\end{figure}

The center region of a UCP was not shielded from all the effects of changes in charge imbalance, however. As the charge imbalance increased, the outer ions expanded faster.  This resulted in the overall volume of the electron gas in the interior increasing faster in turn, which then resulted in a larger amount of electron cooling via expansion. An example of this can be clearly seen in Fig.3, where the high charge imbalance case cooled much more quickly than the low charge imbalance case.

The cooling of the electrons in Fig. 3 is quantified by the decrease in $T$ as a function time. We again note that $T$, while similar, is not identical to the electron temperature. For the $\delta=0.4$ case, the total thermal energy of the UCP electrons is equal to $3/2N_ek_B T$ to an excellent approxmation, as would naively be expected. In contrast, for the $\delta=0.1$ case, the total electron thermal energy is 11\% less than this naive expectation (see Appendix A). However, if both a $\delta=0.1$ and the $\delta=0.4$ UCP with an initial average temperature of $<T_{eff}>$=85K are simulated, a plot of $<T_{eff}>$ vs time would look qualitatively similar to the above figure. The most noteworthy difference would be that the two curves would be slightly closer together, with $\delta$=0.1 curve about 8\% colder at the later times shown in Fig. 3.

The resulting increase in the rate of electron cooling resulted in a small decrease in the expansion rate of the \textit{central portion} of the UCP over the early time evolution of the UCP that we studied. Generally at the very center of the UCP the ions' density distribution is well approximated by a Gaussian over the timescales of the simulation. To characterize the central expansion rate of the UCP, a scale length $\sigma_c$ based on the center ion density was calculated during the expansion. $\sigma_c$ was calculated at each timestep by integrating the ion density from the center outward until a radius within which the number of enclosed ions matched what would be expected to be within one $\sigma$ from the center in a gaussian probability distribution was found. $\sigma_c$ was defined through this radius.

The time evolution of $\sigma_c$ can be seen in Fig. 4. The $\delta$=0.4 simulation  expanded at a slightly slower rate than the center of the $\delta$=0.1 simulation. While the difference between the two expansion rates was small over the course of the simulations, this difference is large enough to surpass the few percent sensitivity of a two cycle rf technique and thus should be experimentally detectable\cite{chen2013}. The fact that increasing $\delta$ leads to overall faster expansion while leading to a slower expansion at the center of the UCP indicates that a change in UCP shape from the initial Gaussian distribution is a fundamental feature associated with changes in charge neutrality.

\begin{figure}
\includegraphics{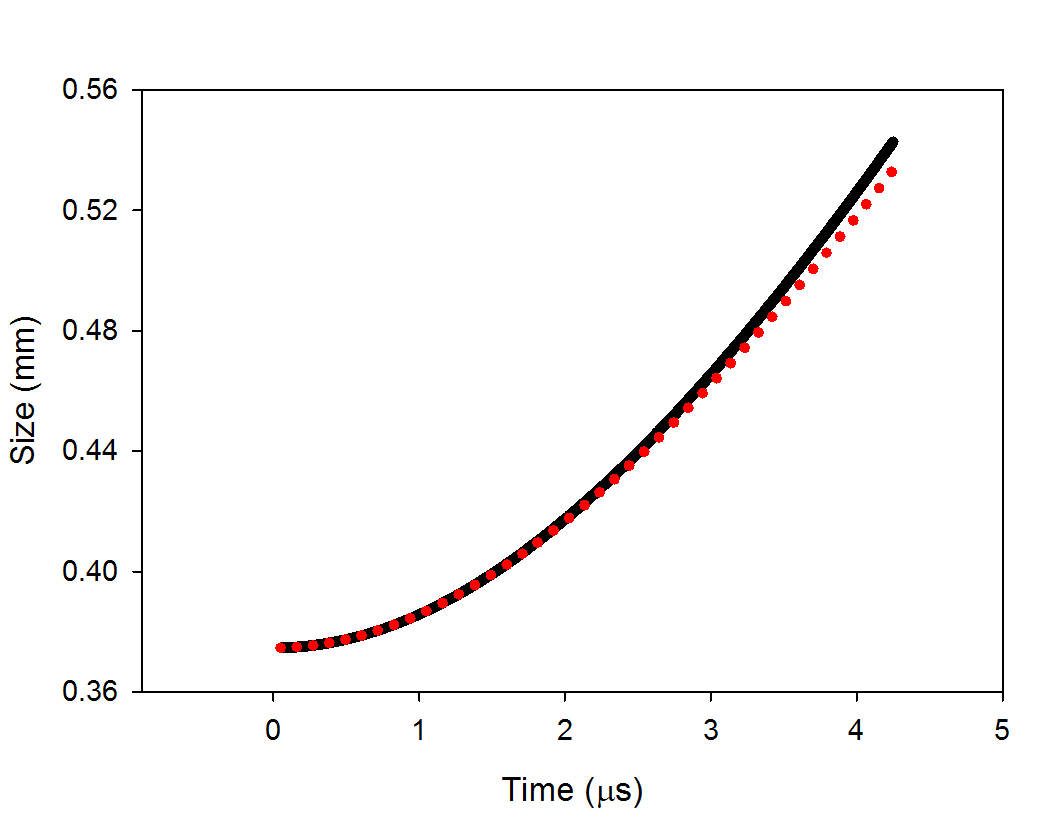}
\caption{Plots of effective center size vs time for two plasmas with different charge imbalances. The solid line is a $\delta$=0.1 plasma, and the dotted line is a $\delta$=0.4 plasma. Both plasma’s have initial parameters $T$=85K, $\sigma_0$=3.75x10\textsuperscript{-4}m, and $N$=3x10\textsuperscript{5}. Since the UCP interior behaves like a neutral plasma, the interior ions can be can be described by a gaussian distribtion at all times.} 
\end{figure} 

The fact that there can be significant impact from the change in charge neutrality on $T$ with only a mild change to the effective size vs. expansion time has experimental implications. Typically, the electron temperature is measured by measuring the rate of change of UCP density\cite{gupta2007,Bergeson2011,Kulin2000}. From this an electron temperature is extracted, often through a computation based on a simple neutral plasma expansion model like Ref\cite{simple} that explicitly assumes perfect neutrality in the interior region of the UCP. For a more non-neutral UCP, this technique would produce an incorrect temperature by not accounting for the interplay between the outer ion expansion and the electron temperature. For a more accurate prediction, a model like the one outlined in this paper would have to be utilized.
 
The robustness of the model results were tested by additional simulation runs over a variety of experimentally achievable initial conditions. This included variations of initial ion number, $T$, and UCP size (i.e. $\sigma_0$). For all tested sets of parameters, UCP expansion and cooling had a qualitatively similar dependence on charge imbalance as the example simulations described above. However, the degree of increased cooling with increasing $\delta$ varied with all of these parameters.

Some care needs to be taken when quantifying how variation in initial parameters affected the additional cooling due to non-neutrality. For example, the initial $T$ of a neutral UCP has a significant impact on the UCP's cooling rate. To properly quantify the  additional cooling caused by non-neutrality as function of initial $T$, it is necessary to decouple the effect that the initial $T$ has on a neutral UCP cooling rate.  This was accomplished by generating cooling curves for $\delta=0.1$ UCPs with different initial values $T$ to serve as a baseline. For each different initial $T$ curve, the times over which the UCP cooled to 75\% of the initial $T$ were recorded. The amount of cooling over these associated time periods was then calculated for $\delta=0.4$ UCPs with the same initial value of $T$. This allowed for a direct comparison of the change in relative cooling rates for two different charge imbalances at a specific initial $T$.

\begin{figure}
\includegraphics{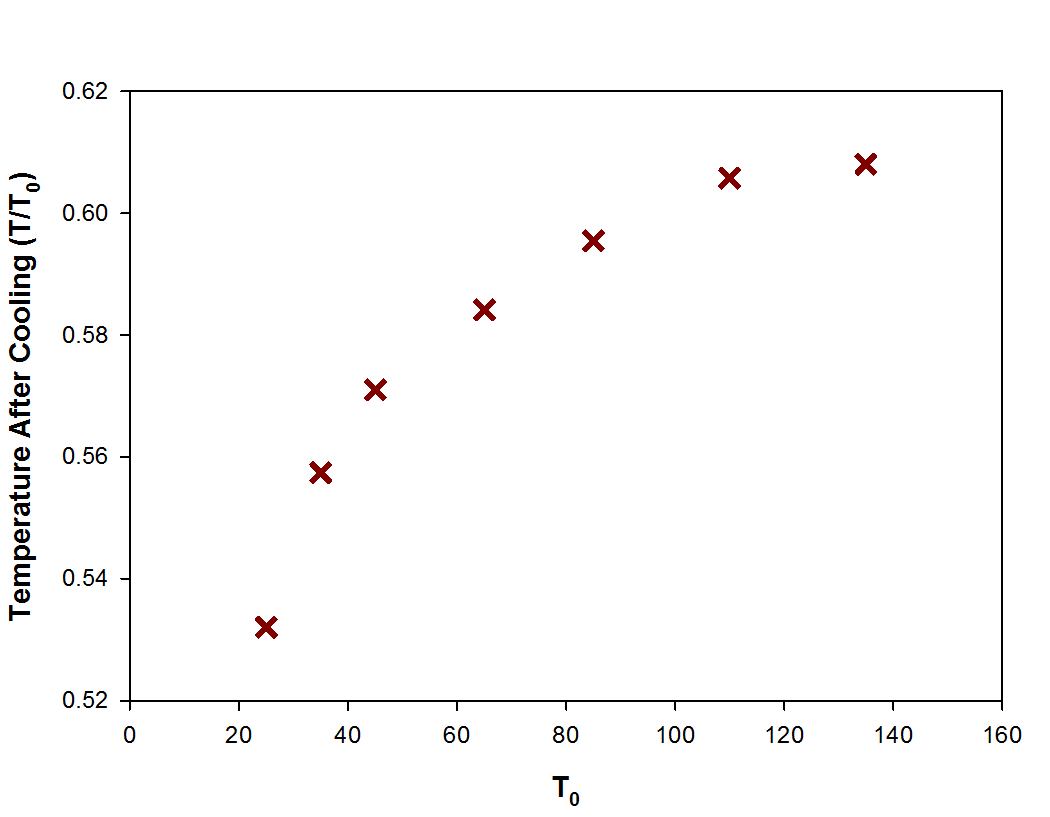}
\caption{The above figure shows the "relative impact" that an increase in charge imbalance had on cooling as a function of initial $T$ ($T_0$). For a given $T_0$, a $\delta$=0.1 UCP was simulated to expand until it had cooled to 75\% of the value of $T_0$. An analogous $\delta$=0.4 UCP was then simulated for the same amount of time as the $\delta$=0.1 UCP. The data points in the figure are the final values of $T$ for the $\delta$=0.4 simulation divided by $T_0$ as a function of $T_0$. The plot shows that the impact on UCP from larger $\delta$, becomes increasingly more important as the initial value of $T$ decreases. All simulations had initial parameters $\sigma_0$=3.75x10\textsuperscript{-4}m and $N$=3x10\textsuperscript{5}. Note that the left axis of this plot does not start at zero.}
\end{figure}

It was found that high $\delta$'s led to more additional fractional cooling in UCPs with low initial values of $T$ than UCPs with high initial values of $T$. This effect can be seen in Fig 5. A similar method revealed that UCPs with higher densities (higher ion number, or smaller size) also experienced more additional charge imbalance induced fractional cooling than UCPs with low densities.
\section{Ion Doppler Profiles}
In addition to investigating the influence of outer ions on the interior region of the UCP, our model can be used to predict the total fluorescence from Doppler-shift sensitive measurements of the ion velocity distribution. Many UCP experiments use an ion absorption imaging to deduce UCP parameters\cite{killian2005,Lyon2011}. Such techniques use near resonant laser pulses to illuminate the UCP. Ion absorption is measured by a CCD camera behind the UCP. By altering the frequency of the laser, the Doppler broadened spectrum for the UCP can be obtained. Mapping the Doppler profiles of the entire UCP yields information on both the size and velocity distributions the UCP ions. For a more in depth discussion of this technique see ref \cite{killian2005}.
\begin{figure}
\includegraphics{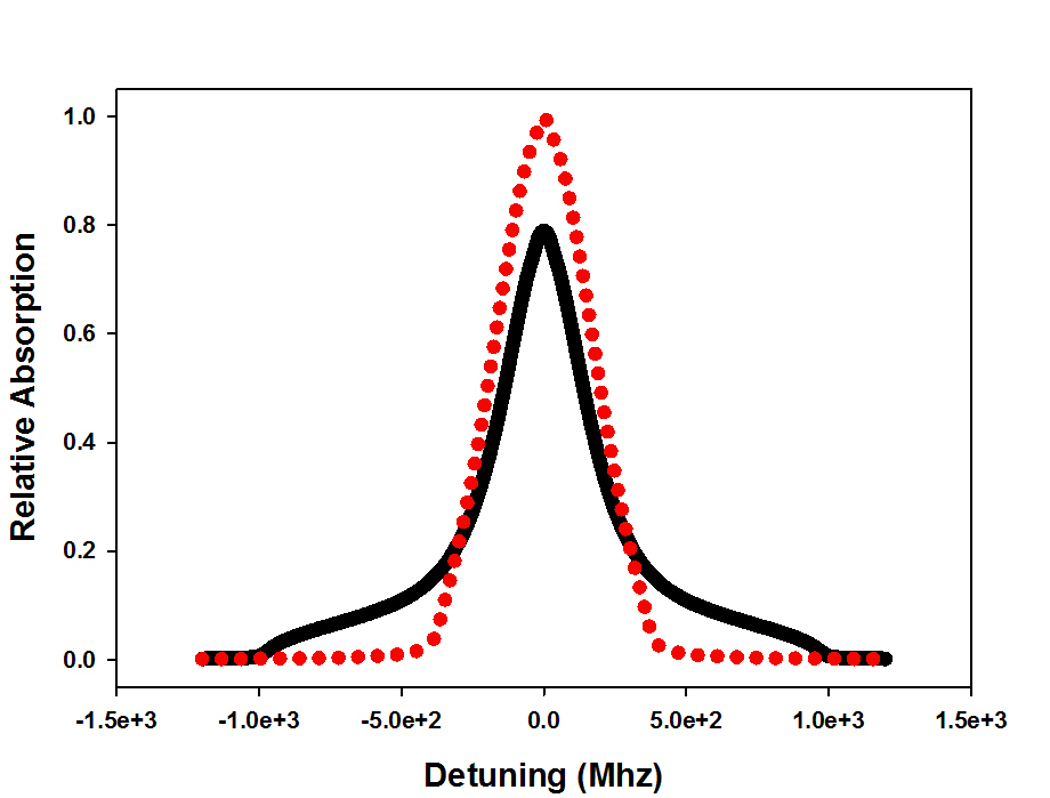}
\caption{Doppler absorption profiles for two plasmas with different charge imbalances at the a t=4.25$\mu$s. The dotted line is a $\delta$=0.1 plasma, and the solid line is a $\delta$=0.4 plasma. Both plasma’s have initial parameters $T$=85K, $\sigma_0$=3.75x10\textsuperscript{-4}m, and $N$=3x10\textsuperscript{5}.}
\end{figure} 

We investigated the effect that non-neutrality would have on such Doppler profiles for for achievable experimental conditions where charge imbalance considerations are more important (e.g. at low density). Such measurements could be used to test the predictions of our model. Conversely, the considerations in this model predict $\delta$-dependent features in such UCP measurements.

To determine if the changes in $\delta$ explored in this work would produce noticeable changes in the Doppler profile, it was necessary to extract theoretical Doppler profiles from model outputs. This was accomplished by calculating the scattering of an incoming uniform intensity laser by the ion shells. The z axis was taken to be the direction of laser propagation. Each ion shell is defined by a single velocity, $v$. However, since $v$ is in the $\hat{r}$ direction, $v_z$ and subsequently the scattering rate changes as a function of x and y position. Thus, it was necessary to integrate across each shell to calculate that shell$'$s scattering rate. The total absorption is calculated by adding up the scattering of each of the shells.  Doppler absorption profiles were determined by calculating the total absorption as function of detuning.

Total absorption profiles were calculated at 4.25 $\mu$s for both the $\delta$=0.1 and the $\delta$=0.4 test cases.  The resulting absorption profiles, seen in Fig. 6, differed significantly. The $\delta$=0.4 profile was broader, most notably in the wings of the distribution, than the $\delta=0.1$ profile. These results indicate that it is important to account for UCP charge imbalance when utilizing ion absorption imaging techniques in situations with reduced charge neutrality.

\section{Conclusion}
We have developed a theoretical model investigating the influence of non-neutrality on the evolution of UCPs. It was found that while the expansion of the UCP interior, over initial expansion times, does not change significantly as a function of UCP neutrality, there is a non-negligible effect on the electron temperature. This effect is predicted to produce an increase in the electron cooling rate  during UCP expansion at higher levels of charge imbalance. A modeling of the effects described in this work are necessary for a proper interpretation of electron cooling experiments in UCPs,  including cooling due to forced and unforced evaporation. Beyond theoretically investigating the influence of charge imbalance on UCP expansion and electron temperature evolution, this model can be extended to include evaporation and three-body recombination induced heating as well to evaluate the likely effectiveness of forced evaporative cooling in achieving greater amounts of electron component strong coupling.  This will be the subject of future work.

\section{Acknowledgements}
We acknowledge support of the Air Force Office of Scientific Research (AFOSR), grant number FA9550-12-1-0222.

\appendix
\section{\\Model Equations} \label{App:AppendixA}
This appendix gives a more detailed description of the equations used in the UCP expansion model described in this work.  The model uses a one demensional scaled potential, $W(R)=\frac{q(\phi(R)-\phi(R_0))}{k_B T}$, where $q$ is electric charge of a proton, $k_B$ is the Boltzmann constant, and $T$ is a model paramenter related to electron temperature which will be further discussed later in the appendix. $R$ is a scaled coordinate, such that $R=r/\sigma_0$ where $r$ is the standard radial coordinate, and $\sigma_0$ is the initial guassian rms parameter for the ion distribution. $R_0$ is boundary condition marking the maximum spatial extent of the electron distribution.

To solve for $W(R)$, Poisson's equation can be written in terms of the scaled potential
\begin{equation}
\frac{1}{R}\frac{d^2}{dR^2}RW(R)=-\frac{N}{N_0}(\Sigma_i(R)-\alpha\Sigma(W(R)))
\end{equation}
where N is the  number of ions, and the scaling constant, $N_0$, is defined as $N_0=4\pi\epsilon_0 k_B T\sigma_0/q^2$. 
The $\Sigma$ and $\Sigma_i(R)$ terms are proportional to the electron and ion charge density, respectively. $\Sigma_i$ is defined by the following the equation:
\begin{equation}
\Sigma_i(R)=4\pi\sigma_0^3 n_i(R)/N
\end{equation}
where $n_i(R)$ is the unscaled ion density determined from the ion shells (see section II). $\Sigma(W)$ is defined as:
\begin{multline}
\Sigma(W)=\\
\begin{cases}
	\sqrt{\frac{\pi}{2}}e^{W}erf(\sqrt{W})-(\frac{2}{3}W+1)\sqrt{2W} & :W> 0 \\
	0 & :W\leq0
\end{cases}
\end{multline}
The $\alpha$ in A1 is a normalization constant defined as:
\begin{equation}
\alpha=\frac{N_e}{N} \frac{1}{\int_0^{R_0}{R^2\Sigma(W(R))dR}}
\end{equation}
where $N_e$ is the total number of electrons.

By simultaneously solving equations A1-A4 and by assuming the boundary conditions $W(R_0)=0$ and $W'(0)=0$, it is possible to obtain a solution for $W(R)$ for all $R\leq R_0$. Once this solution is obtained, it is trivial to determine the electric field applied to the ion cloud.  The field can be expressed as:
\begin{equation}
E=-\frac{k_B T}{q\sigma_0}\frac{dW(R)}{dR} 
\end{equation}
The accelartion of an ion in the $j^{th}$ ion shell, $a_j$, is defined as $\frac{d^2r_j}{dt^2}$, and is equal to the following:
\begin{equation}
a_j=-\frac{k_B T\sigma_0}{m_i}\frac{dW(R)}{dR} \bigg|_{R=r_j/\sigma_0}
\end{equation}
where $m_i$ is the mass of an ion, and $r_j$ is the position of the $j^{th}$ ion shell.

As the UCP expands, the temperature of the electrons cools. As stated earlier, this temperature is not the same as the $T$ in $W$, but the two quantities are closely related. The exact realtion is:
\begin{equation}
\frac{T_{eff}}{T}=\frac{\sqrt{\frac{2}{\pi}}e^Werf(\sqrt{W})-(4/15W^2+2/3W+1)\sqrt{2W}}{\sqrt{\frac{2}{\pi}}e^Werf(\sqrt{W})-(2/3W+1)\sqrt{2W}}
\end{equation}
where $T_{eff}$ is the local electron tempearture for a given W.  For sufficently large W, $T$ is approxiamtely equal to $T_{eff}$. The value of $T$ is determined at all times by using conservation of energy. The total energy of the system, $U$, can be described in these scaled units by:
\begin{multline}
U=2\pi\sigma_0^3\epsilon_0\int_0^{R_0}R^2E^2dR+\frac{1}{2}m_i\sum_{j=1}^{10000}N_j\bigg(\frac{dr_j}{dt}\bigg)^2+\\
\frac{3}{2}k_B\alpha N\int_0^{R_0}R^2\Sigma(R)T_{eff}(R)dR
\end{multline}
where $\epsilon_0$ is the permittivity of free space and $N_j$ is the number of ions in the $j^{th}$ shell.

\bibliography{UCP}

\end{document}